\documentclass[referee]{aa} 
\usepackage{graphicx}

\begin{document}

\title{Broadband microwave sub-second pulsations in an expanding coronal loop
of the 2011~August~10 flare}
\titlerunning {Broadband microwave sub-second pulsations}

\author{ H. M\'esz\'arosov\'a\inst{1}
    \and J. Ryb\'ak\inst{2}
    \and L. Kashapova\inst{3}
    \and P. G{\"o}m{\"o}ry\inst{2}
    \and S. Tokhchukova\inst{4}
    \and I. Myshyakov\inst{3}
     }

\institute{ Astronomical Institute of the Academy of Sciences of the Czech Republic,
            CZ--25165 Ond\v{r}ejov, Czech Republic
       \and Astronomical Institute of the Slovak Academy of Sciences, SK--05960 Tatransk\'a Lomnica, Slovak Republic
       \and Institute of Solar-Terrestrial Physics SB RAS, Irkutsk, Russia
       \and Special Astrophysical Observatory of RAS, St.Petersburg Department, Russia
       }

\offprints{H. M\'esz\'arosov\'a, \email{hana@asu.cas.cz}}

\date{Received ................ / Accepted ...................}

\abstract
{}
{We studied the characteristic physical properties and behavior of broadband microwave
sub-second pulsations observed in an expanding coronal loop during the GOES~C2.4 solar flare on 2011~August~10.}
{The complex microwave dynamic spectrum and the expanding loop images were analyzed with
the~help of SDO/AIA/HMI, RHESSI, and the STEREO/SECCHI-EUVI data processing software, wavelet
analysis methods, the GX~Simulator tool, and the NAFE method.}
{We found sub-second pulsations and other different burst groups in the complex radio
spectrum. The broadband (bandwidth about 1~GHz) sub-second pulsations
(temporal period range 0.07--1.49\,s, no characteristic dominant period) lasted
70\,s in the frequency range 4-7\,GHz. These pulsations were not correlated at their
individual frequencies, had no measurable frequency drift, and~zero
polarization. In these pulsations, we found the signatures of fast sausage
magnetoacoustic waves with the characteristic periods of 0.7 and 2\,s. The other radio
bursts showed their characteristic frequency drifts in the range of
-262--520\,MHz\,s$^{-1}$. They helped us to derive average values of 20--80\,G for
the coronal magnetic field strength in the place of radio emission. It was revealed that the
microwave event belongs to an expanding coronal loop with twisted sub-structures observed
in the 131, 94, and 193\,\AA~SDO/AIA channels. Their slit-time diagrams were compared with
the location of the radio source at 5.7\,GHz to realize that the EUV intensity of the
expanding loop increased just before the radio source triggering. We reveal two EUV
bidirectional flows that are linked with the start time of the loop expansion. Their positions
were close to the radio source and propagated with velocities within a~range of
30--117\,km\,s$^{-1}$.}
{We demonstrate that periodic regime of the electron acceleration in a~model of the
quasi-periodic magnetic reconnection might be able to explain physical
properties and behavior of the sub-second pulsations. The depolarization process
of the microwave emission might be caused by a~plasma turbulence in the radio source.
Finally, the observed EUV flows might be linked with reconnection outflows.}

\maketitle

\section{Introduction}
The fine structures present in the flare radio emissions are studied because these
phenomena might be an effective diagnostics of processes in flare plasma. Of the various
fine structures observed in radio waves (e.g., Ji\v{r}i\v{c}ka et al. 2001), pulsations
have been the subject of many papers, see the reviews, for instance, by Aschwanden (2003, 2004) and
Nakariakov \&~Melnikov (2009).

Generally, radio emissions can be classified into coherent and incoherent emission
mechanisms (Aschwanden 2004). Incoherent emission results from continual processes such as~thermal particle distributions that produce free-free emission (bremsstrahlung) in
microwave and mm wavelengths for low magnetic field strengths
and~mildly relativistic
electron distributions that generate gyrosynchrotron emission,
which is naturally produced during
flares. Coherent emission reflects kinetic instabilities from particle distributions. The
most natural ways to produce these anisotropic particles are~velocity dispersion, which
creates electron beams and thus plasma emission, and~mirroring in~magnetic traps, which
produces loss-cone instabilities (electron-cyclotron emission). The most of coherent
flare-related radio emissions are driven by bursty magnetic reconnection processes and
the associated flare plasma dynamics.

Pulsations with period $P>$~1\,s are frequently observed (e.g., Aschwanden 2003).
Nevertheless, there are also several studies of sub-second pulsations,
for example, in the period
range 0.025--0.055\,s (Xie et al. 2003, Ma et al. 2003), 0.1\,s~(Karlicky et al. 2010),
0.07--0.08\,s (Fleishman et al. 1994), and 0.16--0.18\,s (Fu et al., 1990). There are
rather exceptional observations at microwave frequencies (e.g., in the frequency range
5380--6250\,MHz in Fu et al., 1990).

These sub-second pulsations are typically observed at individual single frequencies, where
we can recognize spikes (Fleishman \&~Melnikov 1998, Fleishman 2004). On the other hand,
we need to see radio dynamic spectra to recognize the~type of the fine structures with the
sub-second pulsating phenomena, for instance, type III pulses in Meshalkina et al. 2012.

Another problem is that radio dynamic spectra observed during solar flares can be very
complex and show different types of bursts and fine structures. One possibility
of studying these fine structures in detail is the separation method (M\'esz\'arosov\'a
et al. 2011a), which is based on the wavelet analysis techniques. This method splits an original
complex (radio) spectrum into two (or more) simpler dynamical spectra according to
the temporal, frequency, and spatial components of individual bursts to simplify the analysis.
This method is suitable when the original radio spectrum~consists of
a~mixture of different fine structures or bursts that are observed at the same frequencies and during
the same time interval, and when it is therefore difficult to recognize individual
temporal or frequency components from each other. The method
also works well~when weak bursts of the spectrum
coincide with strong ones (then we can typically see only the strong component,
while the weak one remains hidden) and~when we wish to detect
or locate possible fast
sausage magnetoacoustic waves.

The properties of the impulsively generated sausage magnetoacoustic waves propagating along
their waveguide (e.g., loops) were theoretically predicted by Roberts et al. (1983,
1984). Each of these fast sausage magnetoacoustic waves form wave trains that propagate
along the waveguide. The time evolution of these trains forms a tadpole wavelet pattern
with a~narrow tail that precedes a~broadband head (Nakariakov et al. 2004). The start of
the wave decay phase corresponds to the tadpole head maximum.

In solar radio observations, these wavelet tadpoles were recognized in the
gyrosynchrotron radio bursts, for example (M\'esz\'arosov\'a et al. 2009a; tadpoles detected at the
same time throughout the whole frequency range), in dm radio fiber bursts generated by the
plasma emission processes (M\'esz\'arosov\'a et al. 2009b, 2011b; slowly drifting tadpoles
corresponding to the frequency drift of the whole group of fiber bursts), and in
sources of narrowband dm radio spikes (Karlick\'y et al. 2011). M\'esz\'arosov\'a et
al. (2013) found these fast magnetoacoustic waves to propagate in the fan structure
of the coronal magnetic null point. These studies were supported by MHD numerical
simulations conducted by Jel\'inek \& Karlick\'y (2012), Pascoe et al. (2013), and
M\'esz\'arosov\'a et al. (2014), for instance.

Thus, the results of the separation method, the analysis of individual observed bursts, and MHD
numerical models for the fast magnetoacoustic wave trains can help us to interpret our
observational data and estimate flare plasma parameters. The goal of our study is to find
a~possible explanation of the broadband microwave sub-second pulsations in a~dynamic
spectrum that was obtained during the 2011~August~10 GOES C2.4 solar flare.

This paper is organized as follows. In Sect.~2 we analyze the observed microwave
dynamic spectrum with broadband sub-second pulsations to determine their characteristic
physical properties. In Sect.~3 we study flare loops belonging to this radio event
with the~help of imaging data (SDO/AIA, SDO/HMI, RHESSI, and
STEREO/SECCHI-EUVI). Finally,
a~discussion and conclusions are presented in Sect.~4.

\section{Analysis of the radio dynamic spectrum with microwave sub-second pulsations}
\begin{figure}[h]
\centering
\includegraphics[width=9.0cm]{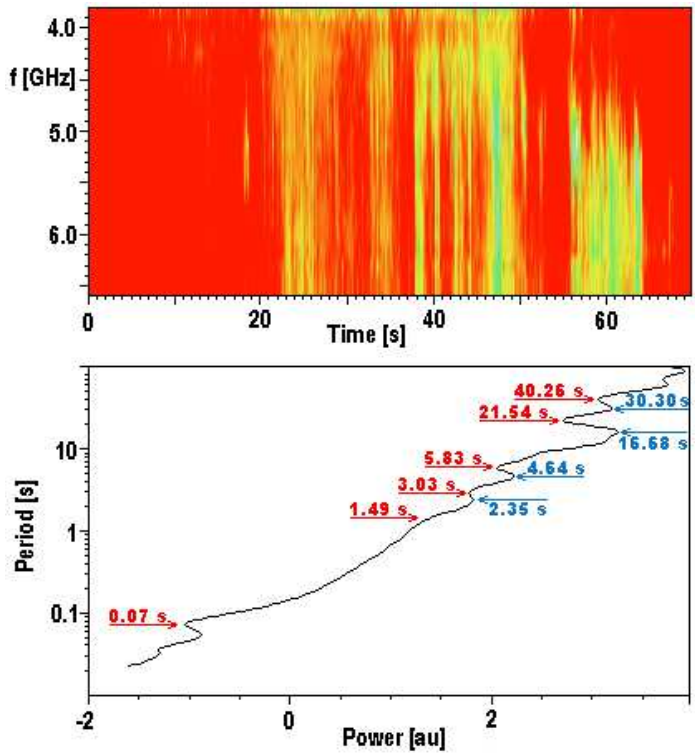}
\caption{Top:~Original complex radio dynamic spectrum with broadband pulsations
that lasted 70\,s (9:33:57--9:35:07\,UT) and was observed by the Badary Broadband Microwave
Spectropolarimeter. Bottom: Averaged global wavelet spectrum~(AGWS) made from the spectrum in
panel~$a)$ with strong peaks at periods~$P$=~2.35, 4.64, 16.68, and 30.30\,s (blue arrows) and with local minima at periods 0.07, 1.49, 3.03, 5.83, 21.54, and 40.26\,s
(red arrows).}
\label{fig01}
\end{figure}

The solar microwave event was observed at 9:33:57--9:35:07\,UT on 2011~August~10 during
the GOES~C2.4 flare that occurred in the active region NOAA~11236. This event was
simultaneously observed with the RATAN-600 (Pariiskii et al. 1976), the Siberian Solar
Radio Telescope (SSRT; Grechnew et al. 2003), and the Badary Broadband Microwave
Spectropolarimeter (BBMS; Zhdanov \& Zandanov 2011, 2015).

The spectral and polarization high-resolution receiver system of the RATAN-600 was
upgraded in 2010, and the current resolution is 0.014\,s (Bogod et al. 2011). This
telescope provides 1D spatial resolution observations that allow us to determine only
the x-coordinate of the flaring source. Generally, the source position can be determined
with the~help of two different projections based on scans measured at different times. With several projections with good angle coverage and resolution, the full map of
a~source can be constructed using this technique (Nindos et al. 1996).

We were unable to determine the position of our sub-second pulsation source with this
technique because the emission was unstable during the time at
which the projections were
taken. We therefore used other properties of RATAN-600 observations. The pulses were recorded by RATAN-600 while
the source was passing through the motionless Gaussian-like beam of RATAN-600 for
about 7\,s with 0.014\,s time resolution. As a~result, we obtained the time profile of
the sub-second pulsations modulated by the RATAN-600 beam profile that is overlaid on a more
stable emitting source. By comparing the RATAN-600 profile with the profile of the sub-second
pulsations from the BBMS instrument, we determined the maximum response of the RATAN-600
beam with respect to the stable emitting source. Then, the~position of this maximum
response was considered as the x-coordinate of the sub-second pulsations. To obtain the
y-coordinate, we used the SSRT data. The result is that the position of the radio source at
5.7\,GHz was at [X,~Y] = [921,~256] arcsec (Kashapova et al. 2013a,~b).

The fine structures including the sub-second pulsations were detected in the radio
dynamic spectrum (Fig.~1, top panel) in the frequency range 3797--8057\,MHz observed by
the BBMS. This radio dynamic spectrum was obtained with a time
resolution of 0.011\,s, and the
event lasted 70\,s (9:33:57--9:35:07\,UT). The BBMS data were measured with~respect to
the left and right circular polarizations (LCP and RCP). We realized that the LCP flux is
equal to that of the RCP, meaning that the microwave event had zero polarization.

\begin{figure*}[ht]
\centering
\includegraphics[width=18.0cm, height=18cm]{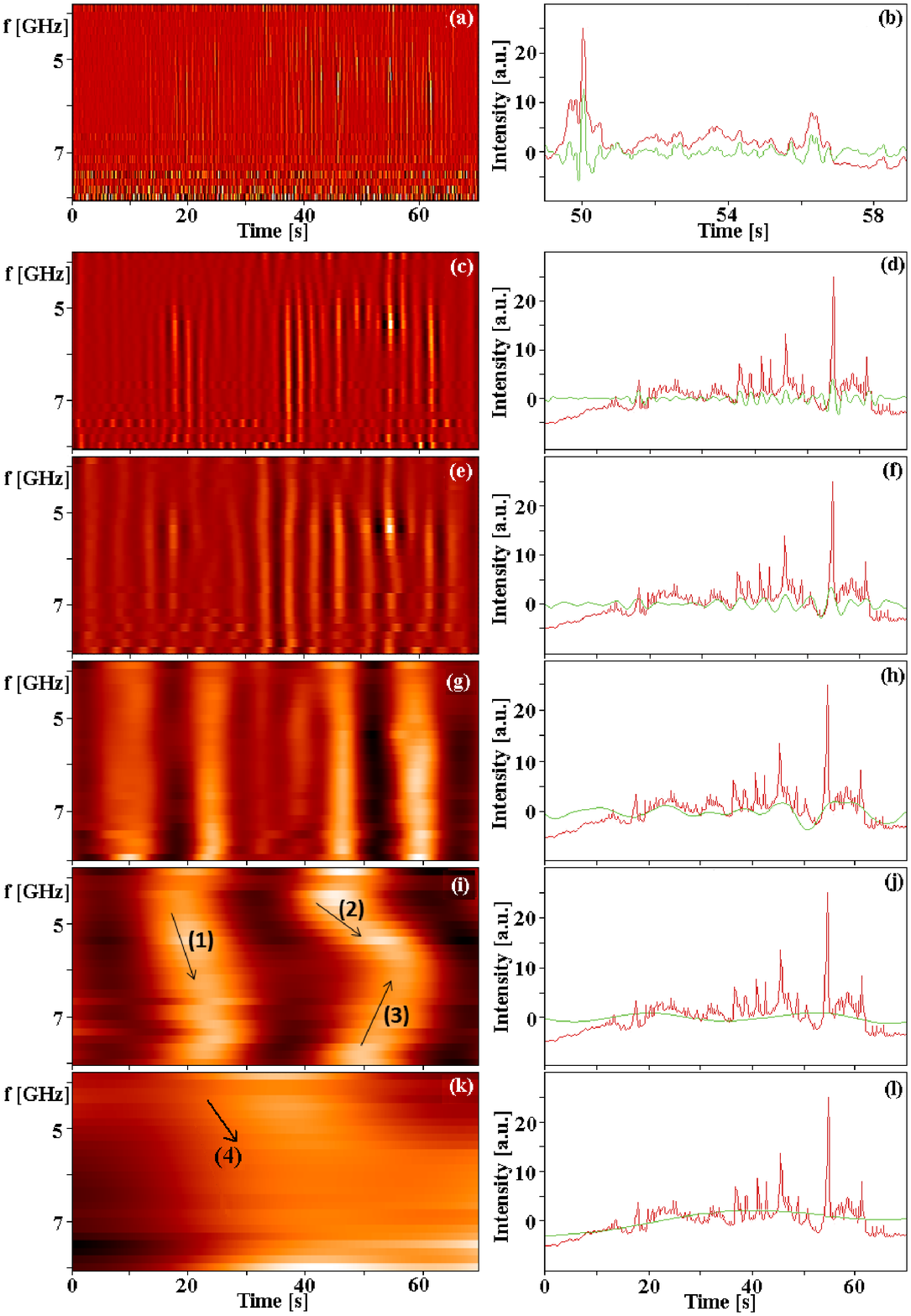}
\caption{Dynamic spectra of separated bursts (left panels) and their selected time series
(right panels). Bursts are separated according to their characteristic period ranges of
0.07-1.49\,s (panel~$a$), 1.49--3.03\,s (panel~$c$), 3.03--5.83\,s (panel~$e$),
5.83--21.54\,s (panel~$g$), 21.54--40.26\,s (panel~$i$), and for periods $>$~40.26\,s
(panel~$k$). The positive and negative parts of the amplitudes (in relation to their mean
values) are given in white and black, respectively (left panels). Frequency drifts of
520, 138, and -262\,MHz\,s$^{-1}$ as well as 158\,MHz\,s$^{-1}$ are marked by arrows in
panel $i$ (1, 2,~and~3) and $k$ (arrow~4). For a~detailed comparison the
individual peaks in the radio time series of the original radio dynamic spectrum (in red) and of the
separated spectrum (in green) are shown for the selected 5194\,MHz in the right panels. To show the
fine individual peaks, we selected the time interval of 49--59\,s (in panel~$b$) within the entire 70\,s
event duration.}
\label{fig02}
\end{figure*}

\subsection{Separation of individual bursts detected in the BBMS radio dynamic spectrum}
We analyzed the BBMS radio dynamic spectrum with~respect to individual types of
observed bursts and their physical parameters. The original complex radio dynamic
spectrum with broadband pulsations is shown in the top panel of Fig.~1 and was observed in
the frequency range 3797--8057\,MHz with a time resolution of 0.011\,s.

To reveal individual burst types that may be included in this radio dynamic spectrum, we
used the separation method (M\'esz\'arosov\'a et al. 2011a) that
is based on the wavelet analysis
technique. We used the temporal periods (for more details see the method
description in M\'esz\'arosov\'a et al. 2011a) to separate the original radio dynamic
spectrum. We computed an average global wavelet spectrum (AWGS) from all time series of
the original radio dynamic spectrum (top panel, Fig.~1). This AWGS curve is shown in
the bottom panel of Fig.~1. This shows individual peaks of the AWGS with
characteristic periods~$P$=~2.35, 4.64, 16.68, and 30.30\,s (blue arrows) as well as
individual local minima with periods of 0.07, 1.49, 3.03, 5.83, 21.54, and 40.26\,s
(red arrows). Then we computed the new separated radio dynamical spectra, each of them
for only a selected period range of the characteristic peak period. For example, for bursts with a~dominant temporal period of~16.68\,s, the new radio dynamic
spectrum was computed in the period range 5.83--21.54\,s, which only shows the bursts of the
selected dominant period (see bottom panel of Fig.~1). The resulting separated spectrum is
displayed in panel~$g$ of Fig.~2.

An overview of all new separated dynamic spectra is presented in Fig.~2 (left panels)
with separation ranges given in the figure caption. Arrows in panels $i$--$k$ show
the frequency drifts of these bursts (see Table~1). The positive and negative parts of
amplitudes (in relation to their mean values) are given in white and black, respectively
(left panels). The period range $<$~0.07\,s contains only instrument interferences (not
presented here). All new separated spectra were computed with the inverse wavelet method
(Torrence~\&~Compo, 1998).

There is no dominant period (no blue arrow in Fig.~1) for the temporal scale
0.07--1.49\,s (panel~$a$, Fig.~2). This is the temporal scale range of the sub-second
pulsations. This means that these pulsations consist of a~mixture of temporal scales where
no individual scale is dominant. Most pulsations displayed in
panel~$a$ of Fig.~2 are broadband (about 1\,GHz).

We computed cross-correlations for individual time series of the dynamic spectrum
displayed in panel~$a$ of Fig.~2. These series are cross-correlated by
50--65\%, but only at the nearest frequencies to each other. This shows that the sub-second
pulsations are not correlated in general.

The bursts we present in the individual left panels (Fig.~2) are not directly visible in the
original radio spectrum (Fig.1, top panel). Therefore, the~validity of these separations is
shown in right panels (Fig.~2). Selected individual time series of the
original radio dynamic spectrum (in red) are compared with the separated one (in green) always at the
same frequency of 5194\,MHz. For example, separated individual peaks
(in green, panel~$b$) belong to the the finest peaks of the original spectrum, that is, to
the sub-second pulsations. The selected time interval in panel~$b)$ is 49--59\,s of the
entire 70\,s event duration to show the finest peaks of the sub-second pulsations. The
separated individual peaks in panels $c$, $e$, and $g$ reflect the different
types of pulses without any measurable frequency drift. Finally, the separated peaks in panels $i$--$k$
show various types of bursts with their significant frequency
drifts (arrows 1--4). Frequency drifts (arrows~1, 2,~and~3 in panel~$i$) are equal to
520, 138, and -262\,MHz\,s$^{-1}$, respectively. The frequency drift (arrow~4 in panel
$k$) is at 158\,MHz\,s$^{-1}$.

\begin{table*}[]
\caption{Characteristic parameters of radio bursts.}
\label{tab1}
\centering
\begin{tabular}{lccccccccccccc}
\hline\hline
F & A & SF & EF  & ST & ET & Dur & FD & S$\rho$ & E$\rho$ & SA & EA & $v_p$ & $B$ \\
  &   &(MHz) & (MHz) & (s)  & (s)  &  (s)  & (MHz\,s$^{-1}$) & (cm$^{-3}$) & (cm$^{-3}$) & (Mm) & (Mm) & (km\,s$^{-1}$) & (G) \\
\hline
2 & 1 & 4408 & 5744 & 17.37 & 19.94 & 2.57 &  520 & 2.41$\times10^{11}$ & 4.09$\times10^{11}$ & 4.24 & 3.39 &  329 &  80\\
2 & 2 & 4408 & 5744 & 41.39 & 51.04 & 9.65 &  138 & 2.41$\times10^{11}$ & 4.09$\times10^{11}$ & 4.24 & 3.39 &   88 &  21\\
2 & 3 & 7607 & 6073 & 44.77 & 50.63 & 5.86 & -262 & 7.18$\times10^{11}$ & 4.57$\times10^{11}$ & 2.68 & 3.24 &   95 &  40\\
2 & 4 & 4519 & 5872 & 24.61 & 33.16 & 8.55 &  158 & 2.53$\times10^{11}$ & 4.28$\times10^{11}$ & 4.15 & 3.33 &   96 &  24\\
4 & 1 & 4887 & 5957 & 61.76 & 62.13 & 0.37 & 2892 & 2.96$\times10^{11}$ & 4.00$\times10^{11}$ & 3.89 & 3.29 & 1610 & 432\\
\hline
\end{tabular}
\tablefoot{F~=~figure number, A~=~arrow number, SF \&~EF =~starting and~ending frequency, ST \&~ET =~starting
and~ending time, Dur =~duration, FD~=~frequency drift, S$\rho$ \&~E$\rho$ =~starting and~ending averaged electron density,
SA \&~EA~=~starting and~ending mean coronal altitudes, $v_p$ = plasma velocity, and $B$ = magnetic field strength.}
\end{table*}

We used the density model of Aschwanden \&~Benz (1995) with numerical values for the height of
$h1$~=~1.6$~\times$~10$^2$\,Mm, for the density of the quiet corona
$n_Q$~=~4.6~$\times$~10$^8$\,cm$^{-3}$, and the parameter $p$~=~2.38. We used this model
to compute characteristic parameters of the individual radio bursts with frequency
drifts in panels $i$ and $k$ of Fig~2. Thus, we obtained the average electron densities at
mean coronal altitudes as well as the plasma velocities.

Then we computed a magnetic field strength $B$~[G]~=~$v_p(S\rho)^{1/2}/2.03\times10^{11}$
for individual drifting bursts, where $v_p$ is plasma velocity [cm\,s$^{-1}$] and $S\rho$
is the starting averaged electron density [cm$^{-3}$]. These characteristic parameters are
shown in Table~1.

These bursts (panels $i$ and $k$, Fig~2) have their positive frequency
drifts in a~range of 138--520\,MHz\,s$^{-1}$ and only one of them shows a negative
frequency drift (-262\,MHz\,s$^{-1}$). While the starting averaged electron density is of
about 2$\times10^{11}$\,cm$^{-3}$ (with one exception for the burst with negative
frequency drift), the ending averaged electron density is always of about
4$\times10^{11}$\,cm$^{-3}$. Similarly, the starting mean coronal altitude is of about
4\,Mm (with one exception for the burst with negative frequency drift), the ending mean
coronal altitudes are always of about 3\,Mm. The plasma velocity is about
90\,km\,s$^{-1}$ and the magnetic field strength is equal to 20--40\,G, except for the burst
(arrow~1, Fig.~2) with the plasma velocity, which equals 329\,km\,s$^{-1}$ and has a magnetic
field strength of~80\,G.

\begin{figure}[h]
\centering
\includegraphics[width=8.5cm]{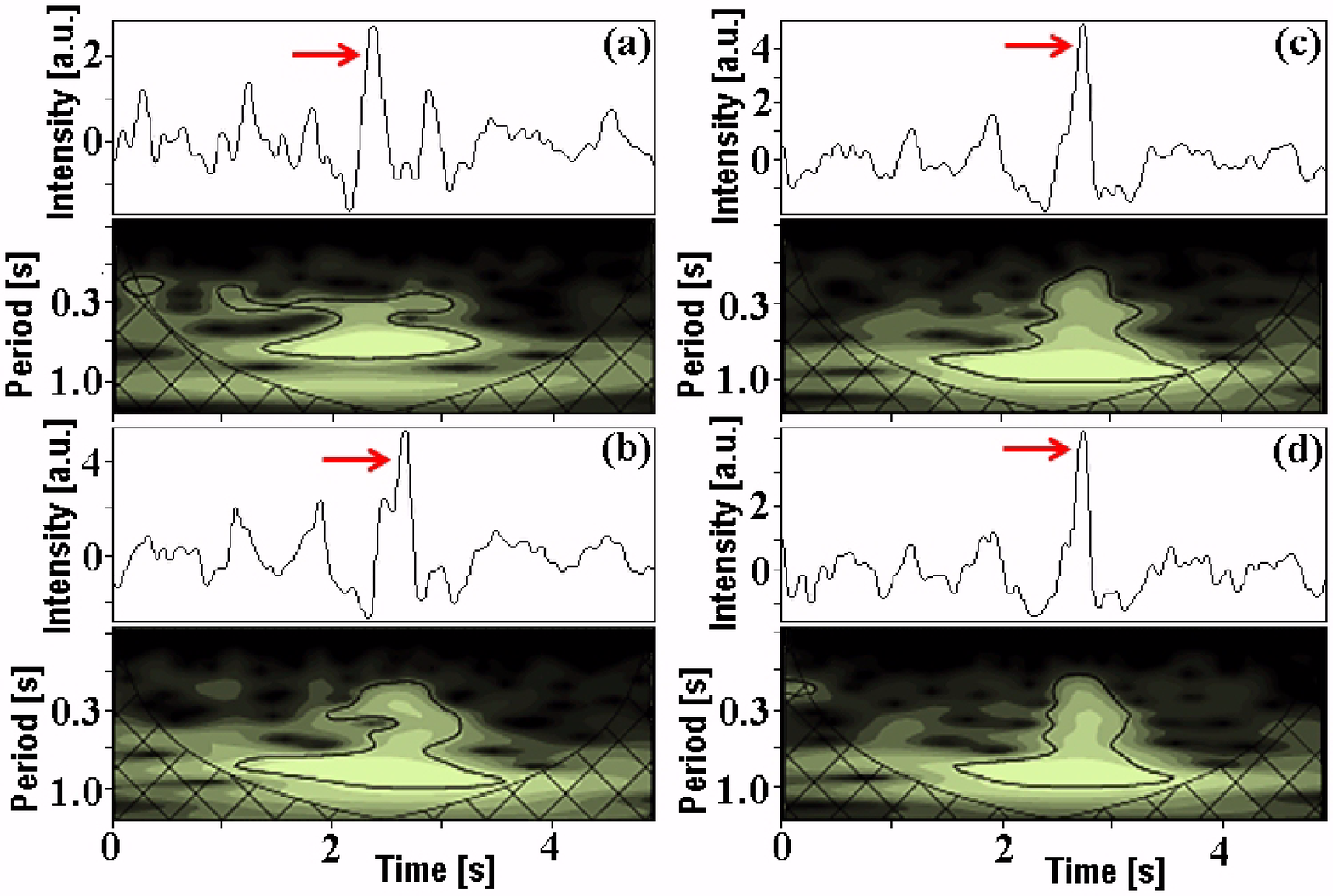}
\caption{Selected time series of the separated radio dynamic spectrum
from panel~$1a$ (Fig.~2) and the wavelet tadpole patterns with a characteristic period
$P\approx$~0.7~\,s. These time series (9:34:56--9:35:01\,UT) observed at frequencies 4887
(panel~$a$), 5579 (panel~$b$), 5957 (panel~$c$), and 6073\,MHz (panel~$d$) show their
peak maxima (red arrows) equal to the maxima of the wavelet tadpole heads. The frequency
drift of these head maxima is 2892\,MHz\,s${-1}$.}
\label{fig03}
\end{figure}

\begin{figure}[h]
\centering
\includegraphics[width=7cm]{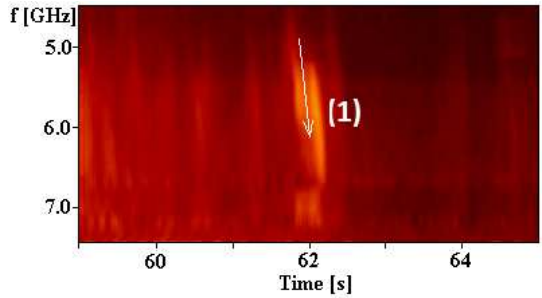}
\caption{Detail (9:34:56--9:35:02\,UT) of the original radio dynamic spectrum (panel~$a$
of Fig.~1) with two close pulses with an average frequency drift
of 2892\,MHz\,s$^{-1}$.}
\label{fig04}
\end{figure}

\begin{figure}[h]
\centering
\includegraphics[width=8.5cm]{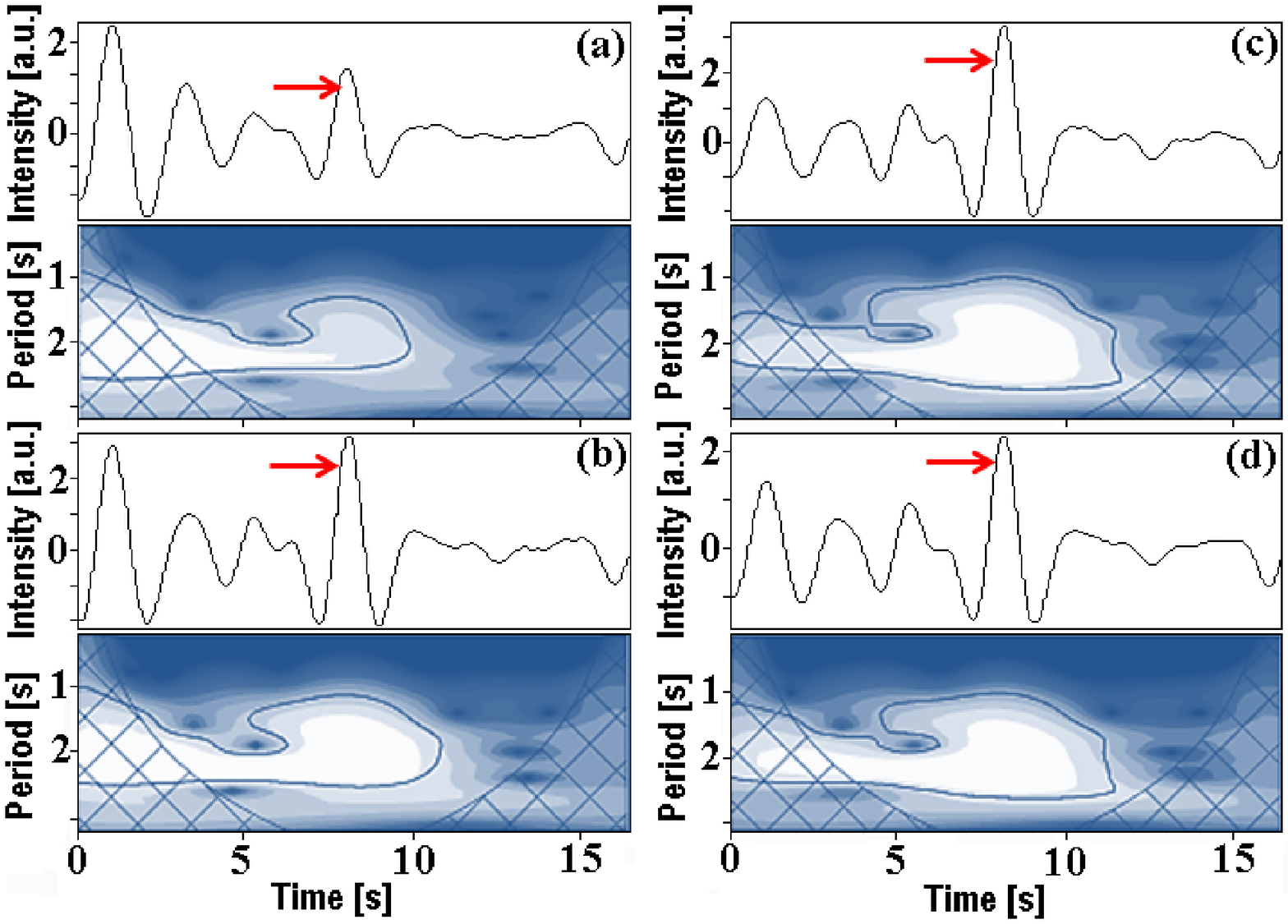}
\caption{Selected time series of separated radio dynamic spectrum in panel~$2a$ (Fig.~2)
and the wavelet tadpole patterns with a characteristic period $P\approx$~2\,s. These time
series (9:34:51--9:35:07\,UT) observed at frequencies 5026 (panel~$a$), 5398 (panel~$b$),
5579 (panel~$c$), and 6593\,MHz (panel~$d$) show their peak maxima (red arrows) equal to
the maxima of the wavelet tadpole heads. There is no measurable frequency drift of these
tadpole head maxima.}
\label{fig05}
\end{figure}

\subsection{Fast sausage magnetoacoustic waves detected in the separated radio dynamic spectra}
We also searched for~possible fast sausage magnetoacoustic waves propagating in situ of
the radio source. It is easier to find them in the time series of the separated
dynamic spectra (Fig.~2) because it is much simpler to find them there than in the original dynamic
spectra (Fig.~1, top panel) using the wavelet separation method (M\'esz\'arosov\'a et al.
2011a).

We found these magnetoacoustic waves in the separated radio dynamic spectra shown in
panels~$a$ and~$c$ (Fig.~2). In Figs. 3~and~5 we present examples of time series and
their wavelet tadpole patterns as signatures of propagating sausage
magnetoacoustic waves.

We show examples of the wavelet tadpoles with the characteristic period
$P\approx$~0.7\,s and the time series in the panels $a$--$d$ of Fig.~3. We selected characteristic time intervals (9:34:56--9:35:01\,UT) of the entire time series
of the separated radio dynamic spectrum (panel~$a$, Fig.~2) where the magnetoacoustic waves
were detected. The selected time series were observed at frequencies 4887 (panel~$a$),
5579 (panel~$b$), 5957 (panel~$c$), and 6073\,MHz (panel~$d$). The series have the peak
maxima (red arrows) at the time of the wavelet tadpole head maxima. The frequency drift of
these head maxima is equal to 2892\,MHz\,s$^{-1}$.

We compared this result with the original time series since the sub-second pulsations have no measurable frequency drift despite of their high bandwidth (about
1\,GHz). We found one exceptional case: Fig.~4 shows a~detail for the time interval of
9:34:56--9:35:02\,UT of the original radio dynamic spectrum with two pulses and with a
frequency drift of 2892\,MHz\,s$^{-1}$, that is, with the same frequency drift as the
wavelet tadpole head maxima in Fig.~3. These pulses occurred at the same frequencies as
the detected magnetoacoustic waves in Fig.~3. The characteristic parameters (Table~1) of
these pulses (arrow~1 in Fig.~4) are significantly different from the other drifting
bursts of Fig.~2 in the frequency drift, event duration, plasma velocity, and magnetic
field strength. On the other hand, such parameters as the starting and~ending averaged
electron density and the starting and~ending mean coronal altitudes are similar for all
observed events (Table~1).

Another case with detected magnetoacoustic waves is presented in Fig.~5. We show
selected time series of the separated radio dynamic spectrum (panel~$c$, Fig.~2) and
their wavelet tadpole patterns with a characteristic period $P\approx$~2\,s. These time
series (9:34:51--9:35:07\,UT) observed at frequencies 5026 (panel~$a$), 5398 (panel~$b$),
5579 (panel~$c$), and 6593\,MHz (panel~$d$) have their peak maxima (red arrows) equal to the
maxima of the wavelet tadpole heads. There is no measurable frequency drift of these
tadpole head maxima in agreement with no f-drift of the bursts in panel~$c$ (Fig.~2).

\section{Analysis of the expanding flare loop in the imaging data}
\begin{figure*}[ht]
\centering
\includegraphics[width=18cm, height=20cm]{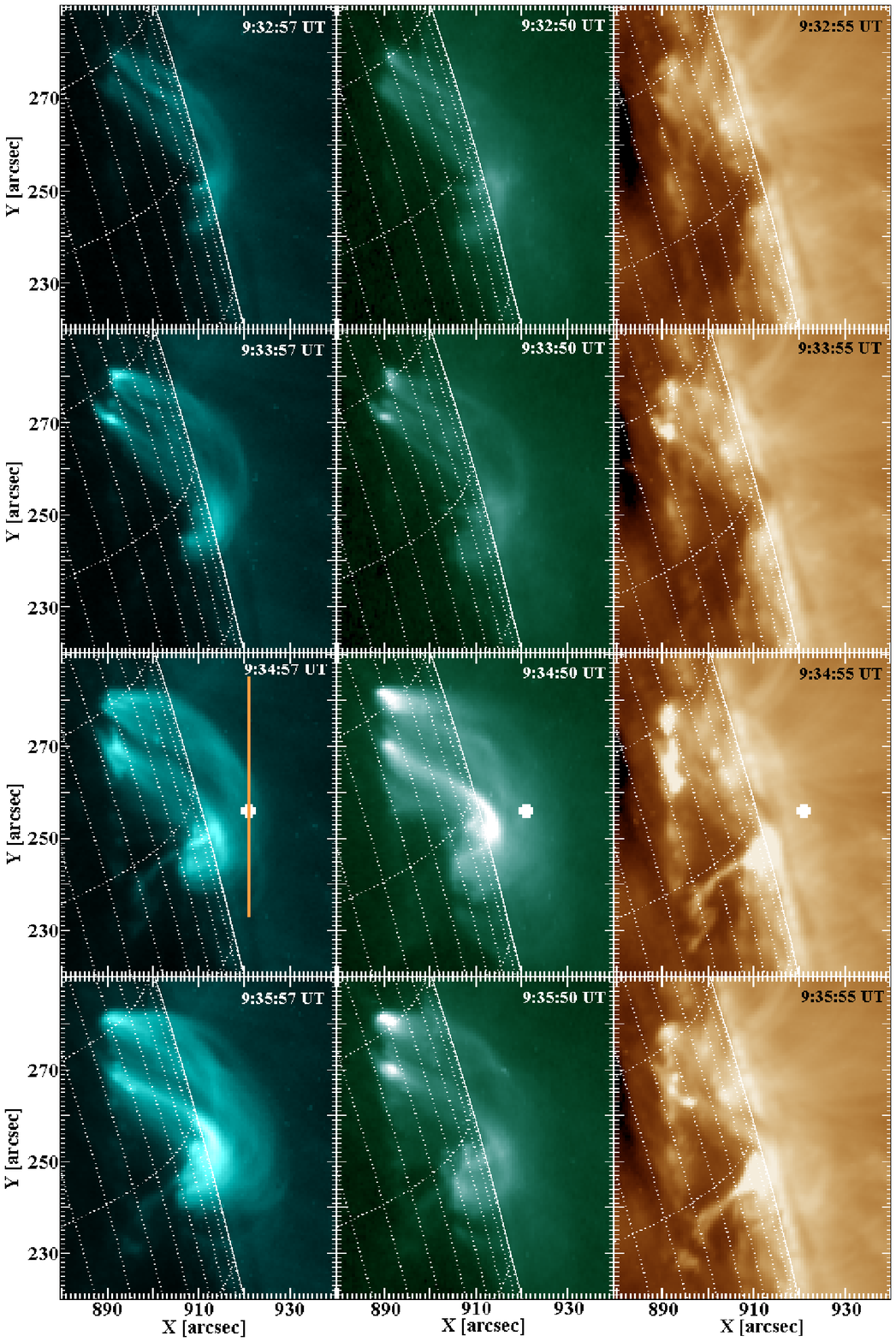}
\caption{Time evolution (9:32--9:35\,UT) of expanding loops observed by SDO/AIA in channels
131\,\AA~(left column), 94\,\AA~(middle column), and 193\,\AA~(right column).
The location of the radio source at 5.7\,GHz is indicated by a~white blob at 9:34\,UT (third
row). An example of the slit position passing the radio source is shown by an orange
vertical solid line in the left column (131\,\AA) for the 9:34:57\,UT panel.}
\label{fig06}
\end{figure*}

To analyze flaring loops and their near environment related to the event we
studied, we used imaging data with high spatial resolution acquired by these instruments:
(i)~EUV images observed by the Atmospheric Imaging Assembly onboard the Solar Dynamics
Observatory (SDO/AIA; Lemen at al. 2012), (ii)~magnetograms observed by the Helioseismic
and Magnetic Imager (SDO/HMI; Schou et al. 2012, Scherrer et al. 2012), (iii) X-ray data
of the Reuven Ramaty High-Energy Solar Spectroscopic Imager (RHESSI; Lin et al. 2002),
and (iv)~EUV images observed by the Extreme Ultraviolet Imager onboard the Solar
TErrestrial RElations Observatory (STEREO/SECCHI-EUVI; Wuelser et~al. 2004). These
imaging data were reduced according to the available instrument manuals using the
standard software tools provided within the SolarSoft package (Freeland and Handy, 1998).

\subsection{Analysis of the SDO/AIA and SDO/HMI imaging data}
We searched for our event in all SDO/AIA channels and found that the event loops
and their footpoints were visible only in the 131, 94, and 193\,\AA~channels, while they cannot be found in the other channels, including the 211 and 335\,\AA~channels. Following the
primary ions and the corresponding temperatures of the individual AIA channels (Table~1 in
Lemen et al. 2012), we can conclude that the emission comes from the hot flare plasma in
the temperature range of characteristic log(T)~=~6.8 (94\,\AA), 7.0 (131\,\AA, Fe~XXI
sensitivity peaks) and 7.3 (193\,\AA, Fe~XXIV ion sensitivity peak) and that there was no
emission for the temperatures below log(T)~=~~6.5 (335\,\AA, 211\,\AA). In the
193\,\AA~channel the loops are only very weakly seen, except
for the loop
footpoints.

The time evolution (9:32--9:35\,UT) of the expanding loops observed by SDO/AIA in channels
131\,\AA~(left column), 94\,\AA~(middle column), and 193\,\AA~(right column) is shown in
Fig.~6. This SDO/AIA event consists of (seemingly) one group of loops at the beginning of
the evolution ($\sim$~9:32\,UT). Later (after $\sim$~9:33\,UT) another
loop is visible that expands toward higher altitudes. This set of two loops (expanded and
non-expanded) shows the highest EUV emissivity of the hot flare plasma (log(T)~=~7.0) at
about 9:34\,UT and later (see Fig.~7).

The location of the radio source at 5.7\,GHz is indicated by a~white blob at 9:34\,UT
(third row in Fig.~6). This shows that the radio source (studied in Sect.~2) was observed at the time
of the increased EUV emission visible in the 131 and 94\,\AA~and
when it was localized in the
expanding loop.

We studied the 131, 94, and 193\,\AA~data in more detail with the~help of time series
(temporal evolution during interval 9:31:33--9:39.57\,UT) derived for the individual
vertical slits of the fix position in the axis~Y (233--286\,arcsec) and of different
X-axis~ positions in the range of 881--928\,arcsec with a~step of 1\,arcsec. An example of the
slit position passing the radio source is presented by the orange vertical solid line in the
left column (131\,\AA) for 9:34:57\,UT panel (Fig.~6).

\begin{figure}[h]
\centering
\includegraphics[width=8cm]{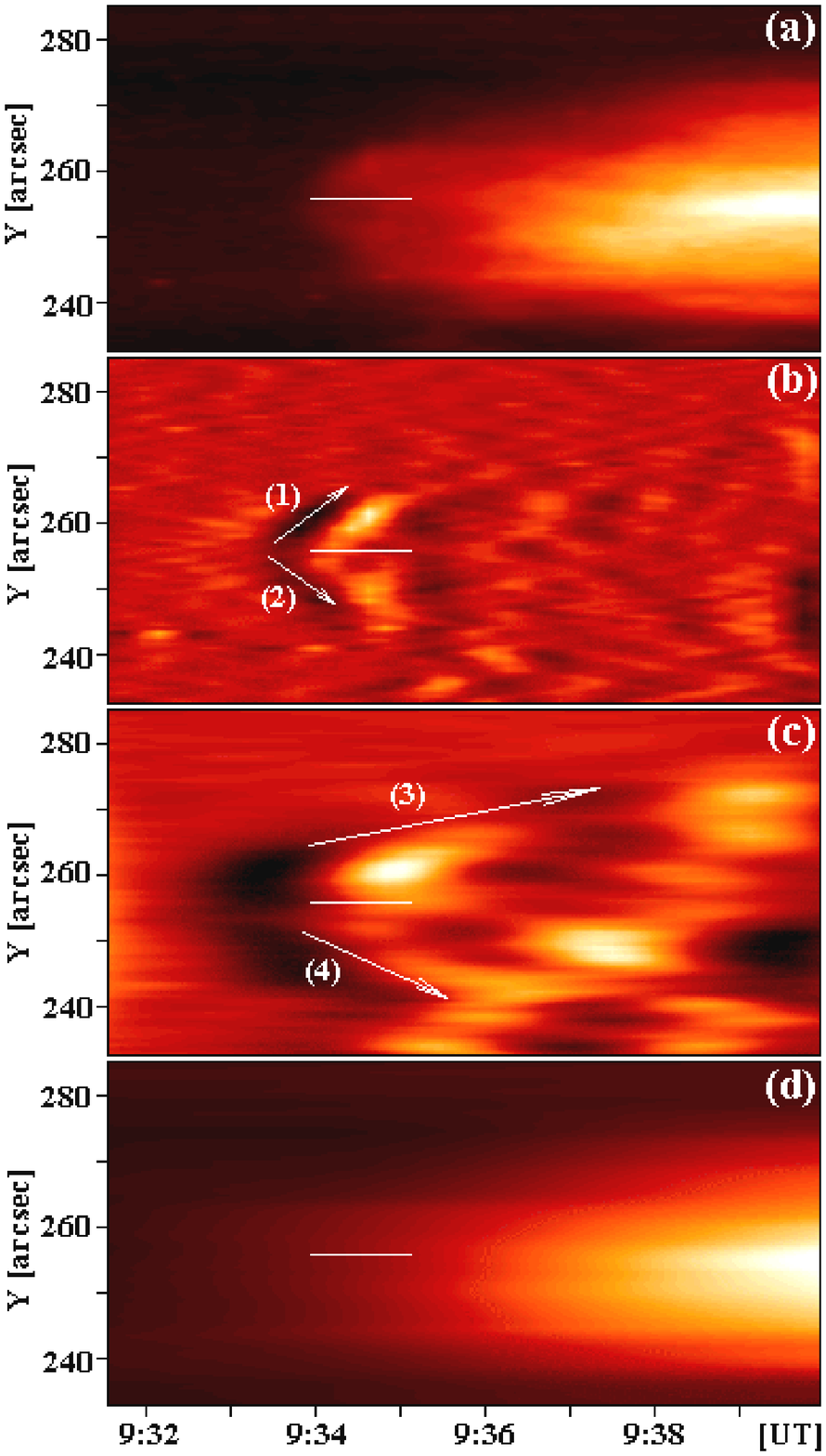}
\caption{Example of time-slit diagram (9:31:33--9:39:57\,UT) for the SDO/AIA~131\,\AA~channel
and slit at position X~=~921\,arcsec / Y~=~233--286\,arcsec. Panel~$a$: Original
time-slice diagram. Separated time structures according to their characteristic temporal
scales 42--120\,s, 120--348\,s, and $>$~348\,s are shown in panels $b$,~$c$, and~$d$,
respectively. The position (Y~=~256\,arcsec) and radio source time interval
(9:33:57--9:35:07\,UT) is presented by the horizontal solid line in individual panels.
Individual arrows mark detected EUV flows propagating in the
loop area. Estimated
velocities are 117, 109, 30, and 70\,km\,s$^{-1}$ for arrows 1,~2,~3, and~4,
respectively.}
\label{fig07}
\end{figure}

An example of the time-slit diagram for the SDO/AIA 131\,\AA~channel for the time
interval 9:31:33--9:39.57\,UT and the slit at the position X~=~921\,arcsec and
Y~=~233--286\,arcsec is presented in Fig.~7. Panel~$a)$ shows the original time-slice
diagram with the EUV emission that started at a time of about 9:34\,UT at acoordinate about
Y~=~255\,arcsec. The highest intensity is displayed after 9:37\,UT at
Y~$\approx$~250--265\,arcsec.

The position and temporal extent (9:33:57--9:35:07\,UT) of the radio source, observed
at 5.7\,GHz, is shown with the~horizontal solid line in the individual panels of Fig.~7.

The individual time-slit temporal structures in the AIA~131\,\AA~channel emission were
separated according to their characteristic temporal scales using the wavelet separation
method of M\'esz\'arosov\'a et al. (2011a). The separated emissions are shown in Fig.~7. Panels~$b)$, $c)$, and~$d)$ show separated temporal structures revealed at their
characteristic temporal scales of 42--120\,s, 120--348\,s, and scales $>$~348\,s,
respectively. Different 131\,\AA~flows are shown in panels $b)$~and~$c)$. These
bidirectional flows start before the radio emission begins ($\approx$~9:33\,UT)
and are located at the Y-coordinate of the radio source (Y~=~256\,arcsec). Then the flows
propagate toward the coordinates Y~=~280 and Y~=~240\,arcsec. The individual arrows
1,~2,~3, and~4 (Fig.~7) show different plane-of-sky flow velocities along the loop with values of 117, 109, 30, and 70\,km\,s$^{-1}$, respectively. The flow displayed in
panel~$b)$ seems to be connected mainly with the radio source (in time as well as in
place). The flows in panel $c)$ might be connected with the radio source as well as the
loops footpoints. Panel~$d)$ shows the 131\,\AA~emission, which starts about the time and in the approximate location
of the radio source and propagates in a~similar way as the flows in panel~$c)$. The
highest emission intensity is detected after 9:38\,UT and later (beyond the~scope of our
study).

The time-slice diagrams at different slit positions around X~=~921\,arcsec are similar to the
time-slice diagram presented in Fig.~7. Similar results were also obtained for the
94\,\AA~channel. We detected no kink oscillations in our time-slice diagrams.

\begin{figure}[h]
\centering
\includegraphics[width=9cm]{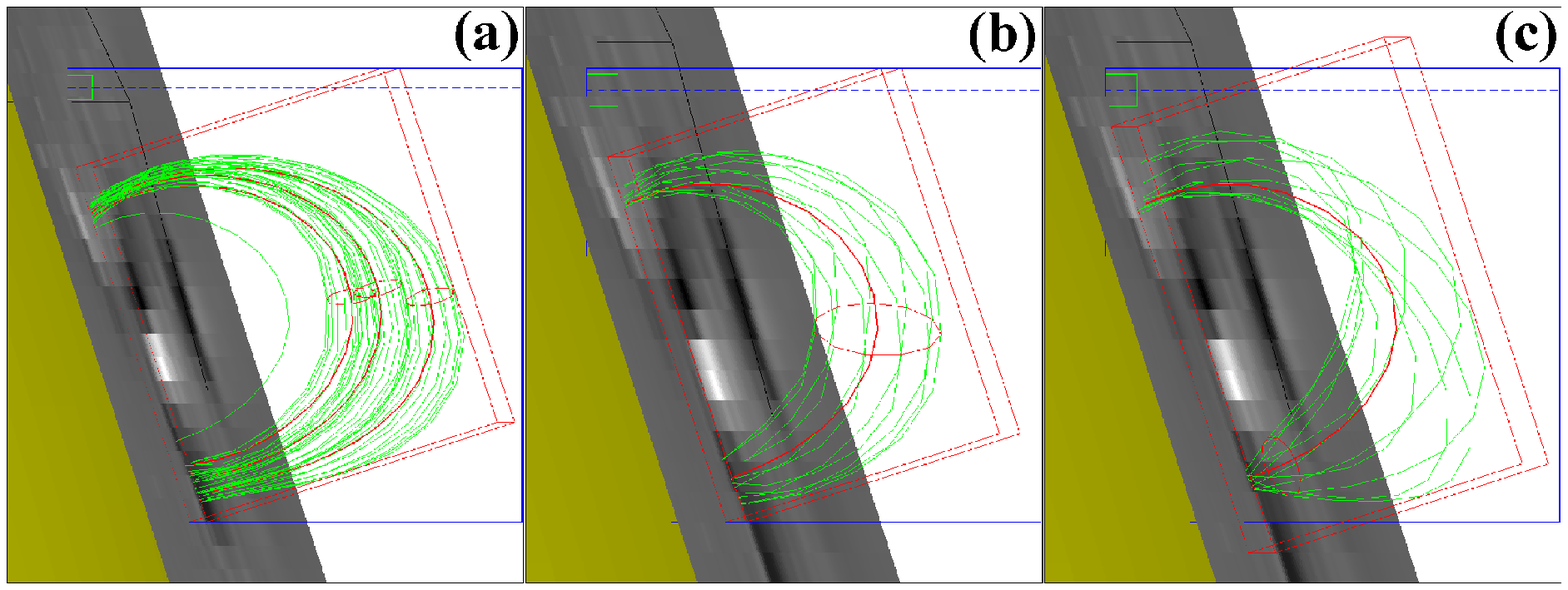}
\caption{Three-dimensional two-loop model of the magnetic field line configuration obtained from
a potential extrapolation in the $GX~simulator$. The panels show reconstructed
magnetic field lines of 20--40\,G (panel~$a$), 80\,G (panel~$b$), and 450~G (panel~$c$).}
\label{fig08}
\end{figure}

\begin{figure}[ht]
\centering
\includegraphics[width=9cm, height=17cm]{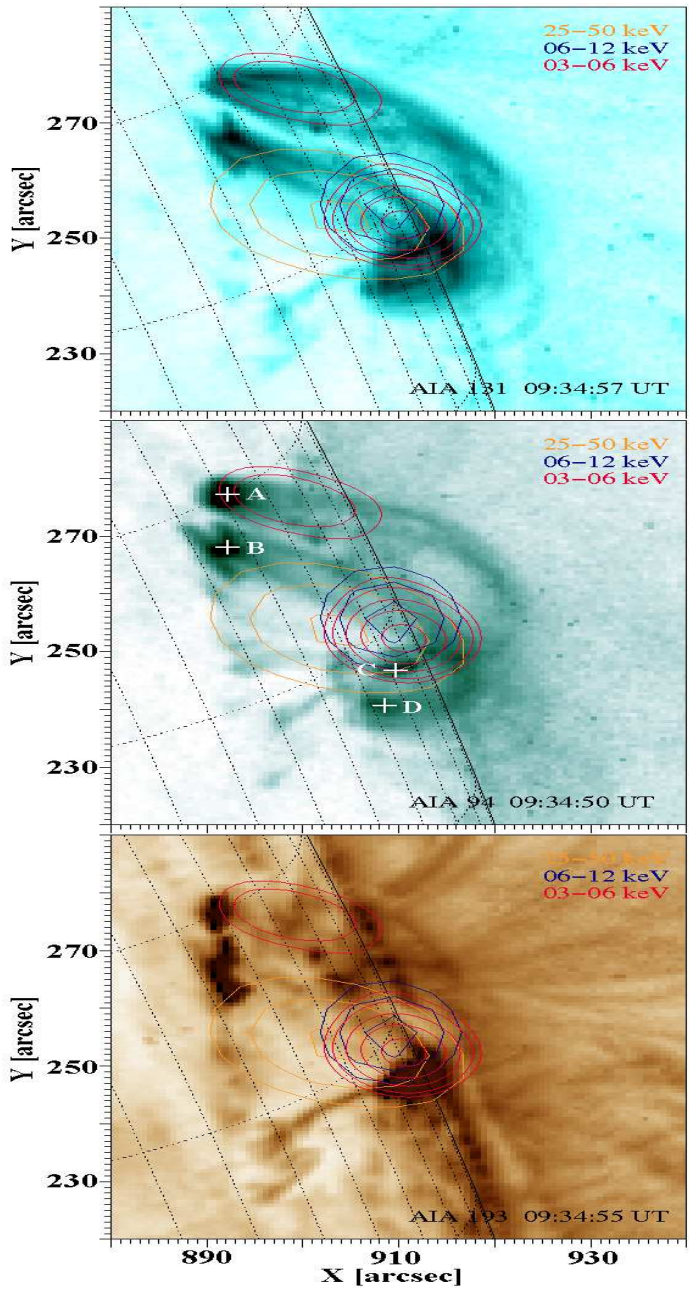}
\caption{Individual loops observed by SDO/AIA in channels 131, 94, and 193\,\AA~at
9:34\,UT processed with the NAFE method. Each panel is overlaid by RHESSI X-ray sources in
different energy bands shown as contours: 03--06\,keV (in red), 06--12\,keV (in blue),
and 25--50\,keV (in yellow). The integration time was 9:34:00--9:35:00\,UT for the
3--6\,keV band and 9:34:20--9:35:00\,UT for the 6--12 and 25--50\,keV bands. Individual
contours are displayed for 65, 82, and 98\% (25--50\,keV band), 79, 87, and 95\%
(6--12\,keV band), and for 47, 56, 69, 82, and 95\% (3--6\,keV band) of maxima in their
intensity. Loop footpoints are marked by crosses A--D in middle panel.}
\label{fig09}
\end{figure}

\begin{figure}[ht]
\centering
\includegraphics[width=9cm, height=18cm]{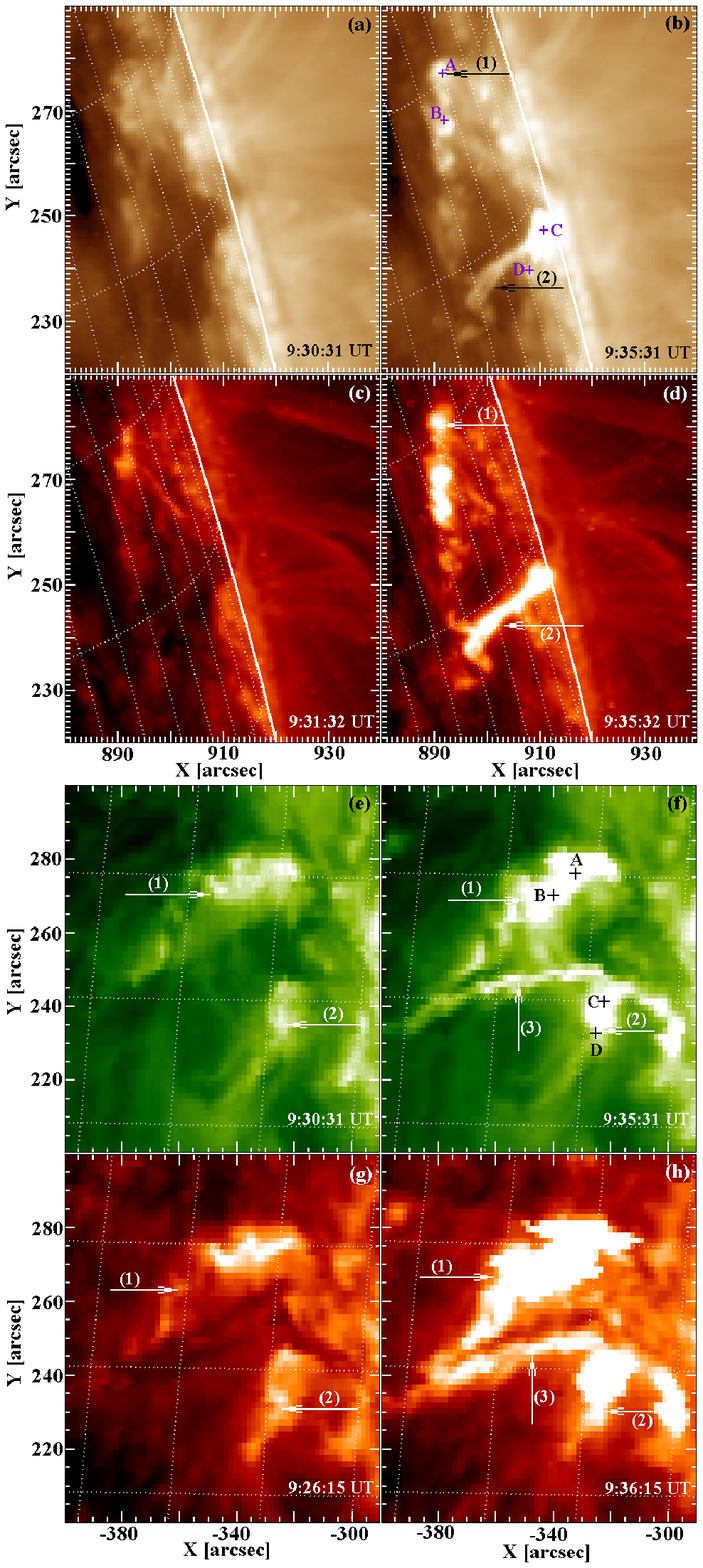}
\caption{Time evolution of individual flare structures (footpoint ribbons)
observed by SDO/AIA (panels $a$--$d$) and STEREO/SECCHI-EUVI (panels $e$--$h$).
Observations are provided at channels 193\,\AA~(panels $a$--$b$), 304\,\AA~(panels
$c$--$d$), 195\,\AA~(panels $e$--$f$), and 304\,\AA~(panels $g$--$h$). Panels in the left and
right columns reflect the situation before (9:26--9:31\,UT) and after (9:35--9:36\,UT)
formation of the new flare structure (arrow~3) that is visible between the older ones marked by
arrows 1~and~2. Positions of individual loop footpoints (see middle panel in Fig.~9)
are marked by crosses A--D in panels~$b$ and~$f$.}
\label{fig10}
\end{figure}

We also used SDO/HMI data because of the situation of the magnetic field lines of flaring loops
with respect to their characteristic values of magnetic field strength. These SDO/HMI
data were processed using the $GX~Simulator$ tool (Nita et al. 2015) to reconstruct
a~three-dimensional two-loop model of the magnetic field line configuration. This model was
obtained from a~potential extrapolation based on the SDO/HMI data of the event. The panels in Fig.~8 show reconstructed approximate magnetic field lines
of 20--40\,G (panel~$a$), 80\,G (panel~$b$), and 450~G (panel~$c$). While these magnetic
field lines of a~weak magnetic field are rather parallel, dense, and concentrated mainly
around the expanded loop (panel~$a$), the magnetic field lines of the stronger magnetic
field are rather less structured and more dispersed throughout the entire active area
(panels~$b$--$c$).

We used the magnetic field extrapolation in the potential approach in our study
because the limb effects are strong in the reconstructed region. This is not a
fully correct description of the magnetic field characteristics during the flare. Several indicators make this suitable, however: (1)~the magnetic field lines shown
in Fig.~8 and the flare loops seen in EUV (Fig.~6) agree well. (2)~The
reconstructed magnetic field lines obtained by the nonlinear force-free approach using
an optimization method (Rudenko \& Myshyakov, 2009) also agree
well, but in
this case we cannot see fine details (twisted magnetic field lines) because of the limb
effects. (3)~The results agree well with the theoretical modelling for
a gyrosynchrotron model of the radio emission, for instance (Kashapova et al 2013a).

\subsection{RHESSI and STEREO/SECCHI-EUVI imaging data of the loops}
We compared RHESSI and STEREO/SECCHI-EUVI observations with the SDO/AIA data to
obtain more information about the active area.

However, as a first step, the individual loops observed by SDO/AIA in channels 131, 94, and
193\,\AA~at 9:34\,UT were processed by the noise adaptive fuzzy equalization method
(NAFE; Druckm{\"u}ller 2013), and the result is shown in Fig.~9. The NAFE method allows us to
see the individual loop sub-structures better (compare the panels of
Fig.~9 with those in the third row in Fig.~6). Thus, the panel for
131\,\AA~(Fig.~9) shows that the expanding loop might consist of about three twisted individual
sub-structures that possibly cross each other. The non-expanding loop possibly consists of two
parallel loops in~reality. Positions of the main loop footpoints are marked by crosses
A~--~D in the middle panel. It seems that the expanding and non-expanding loops are
placed between the footpoints A~and~C and B~and~D, respectively.

Each panel in Fig.~9 is overlaid by RHESSI X-ray sources in the 3--6\,keV band,
6--12\,keV band, and 25--50\,keV band. All panels of Fig.~9 show
that the
highest intensity of all X-ray bands is focused at the loop footpoints and below the
expanding loop.

We compared the SDO/AIA and STEREO/SECCHI-EUVI observations. The time evolution
of the individual observed flare structures (flare ribbons) is shown in Fig.~10. The SDO/AIA (panels $a$--$d$) and STEREO/SECCHI-EUVI (panels $e$--$h$) observations are in
the channels 193\,\AA~(panels $a$--$b$), 304\,\AA~(panels $c$--$d$), 195\,\AA~(panels
$e$--$f$), and 304\,\AA~(panels $g$--$h$). The panels in the left column reflect the situation
before the flare (9:26--9:31\,UT), and the panels in the right column the situation after the
flare (9:35--9:36\,UT). The formation of a~new flare structure visible between the older
ones (arrows 1~and~2) is marked by arrow~3.

To identify these particular footpoint positions detected in the SDO/AIA
images (Fig.~9) in the STEREO/SECCHI-EUVI images (Fig.~10) we used transformations based
on the coordinate system definitions for solar images (Thompson, 2006), the dedicated
routines (see, e.g., Thompson \& Wei, 2010) from the SolarSoft package (Freeland \& Handy,
1998), and information stored in the FITS data file headers. In particular, images taken
at 09:39:50 UT and 09:35:30 UT by the SDO/AIA and STEREO/SECCHI-EUVI instruments were
used. We computed the positions of footpoints A--D for different values of
radial heights in a~range 0--5000 km with a~step of 100\,km. In all these values we
searched for those that might be connected with the postflare ribbons.

The resulting footpoint positions in the STEREO/SECCHI-EUVI images are shown in panel~$f$
of Fig.~10. The positions of footpoints A, B, C,
and~D are indicated by crosses at the radial heights of 1.4, 1.5, 3.5, and 2.3\,Mm, respectively. The footpoints
computed for other values of the radial heights were outside of the ribbons under study.
Footpoints A~and~B are connected with the~part of the
ribbon (coordinate~Y~$\approx$~270--280\,arcsec) where higher EUV intensity was observed.
Moreover, these footpoints were detected with a~similar radial height (1.4 and 1.5\,Mm).
Footpoint~C is connected with the part of the ribbon
(coordinate~Y~$\approx$~240\,arcsec) with the highest EUV intensity observed by both the
SDO/AIA and the STEREO/SECCHI-EUVI instruments. Footpoint~D is localized below this
part of the ribbon (coordinate~Y~$\approx$~230\,arcsec).

\begin{figure}[ht]
\centering
\includegraphics[width=5cm]{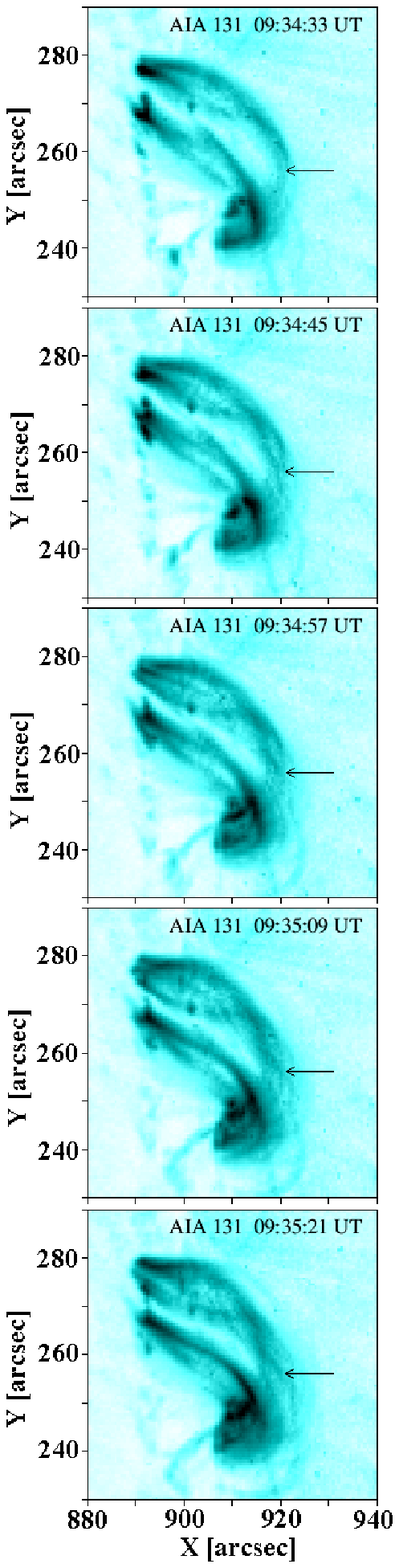}
\caption{Fine sub-structures of the expanding loop observed by SDO/AIA
in 131\,\AA~channel during 9:34:33--9:35:21\,UT processed with
the NAFE method. The position of the
radio source is marked by the arrow.}
\label{fig11}
\end{figure}

\section{Discussion and conclusions}
We studied the solar microwave event observed at 9:33:57--9:35:07\,UT by the Badary Broadband
Microwave Spectropolarimeter on 2011~August~10 during the GOES~C2.4 flare, which occurred in the
active region NOAA~11236. The position of the background flaring radio source at 5.7\,GHz was
obtained using the SSRT and RATAN-600 observation at this radio frequency (Kashapova
et al. 2013a,~b).

We revealed that the microwave radio dynamic spectrum (Fig.~1, top panel) observed in the
frequency range (3797--8057\,MHz) with zero polarization consists of broadband (about
1\,GHz) sub-second pulsations and other different bursts (Fig.~2). Some bursts have frequency drifts in the range -262--2892\,MHz\,s$^{-1}$ (Table~1). We derived
values of the magnetic field strength in the range 21--432\,G for these bursts with the frequency
drift.

The signatures (wavelet tadpole patterns) of fast sausage magnetoacoustic waves
propagating in situ of the radio source were detected in our data. In the
first case (Fig.~3), the wave characteristic period is 0.7\,s and the frequency drift of
the tadpole head maxima is 2892\,MHz\,s$^{-1}$ , which is the same as the frequency drift of
the pulses (Fig.~4) observed in the same time as the magnetoacoustic waves. In the second
case (Fig.~5), the wave characteristic period is about 2\,s and there is no frequency
drift of the tadpole head maxima.

According to the SDO/AIA 131 and 94\,\AA~images, only two seemingly
simple and faint loops were observed near to the limb before 9:32:21\,UT. From this moment,
one of the loops (on the higher Y-coordinate values) expanded toward higher altitudes. The
second loop remained more or less at its original position. The expanding loop spread to
an increasingly greater area, and new loop sub-structures were created here eventually. This raises the~question why just this one loop of the two expanded. A~possible reason for
this situation can be seen in the STEREO/SECCHI-EUVI top point of view of the active region shown in
Fig.~10. This shows the situation (left column) of two very weak footpoint ribbons
that are indicated with arrows~1 and~2. In the right column the later situation is shown, when both
ribbons were linked with a~new flare structure marked by arrow~3. From this moment on, both
ribbons (arrow~1 and~2) became highly excited and influenced the loop
footpoint area. In this area the highest RHESSI X-ray source intensity (Fig.~9) was
focused. It seems that the two footpoints of the expanding loop were excited at
about 9:32:21\,UT.

The SDO/AIA 131\,\AA~emission intensity increased from about 9:32:21\,UT, as shown by the
time-slit diagram in Fig.~7. The radio source was triggered early after
the start of the loop expansion, at 9:33:57\,UT (horizontal solid line in Fig.~7).
The earlier EUV flows are presented in panel $c)$. This bidirectional perturbation was
generated with the start time of the loop expansion (9:32:21\,UT) and propagated along the
slit (vertical orange solid line in the left panel of Fig.~6) with a~velocity of 30 and
70\,km\,s$^{-1}$ toward the higher altitudes and the loop footpoint, respectively. Thus,
the flow propagation in panel~$c)$ is significantly faster in the direction toward the
footpoint.

The later EUV flows (Fig.~7, panel~$b$) might be linked with triggering the radio source
as well as with the generation of sausage magnetoacoustic waves. This bidirectional
perturbation was generated just before the start time of the radio source emission
(9:33:57\,UT) and propagated with a~velocity of 117 and 109\,km\,s$^{-1}$ toward the
higher altitudes and the loop footpoint, respectively. These perturbation velocities
(arrows~1 and~2) are similar to each other, and the propagation of these flows (in
panel~$b)$ is significantly faster in both directions that the earlier flows (in
panel~$c$). While the earlier perturbation propagated over the
entire area of the expanding loop
(Y~=~280--230\,arcsec), the later one was instead focused around the radio source.

The main subject of our study was the origin of the broadband (bandwidth about 1\,GHz)
sub-second pulsations (temporal period range 0.07--1.49\,s, no dominant period) that lasted
70\,s in microwaves (about 4--7\,GHz). Except for one (Fig.~4), these pulsations
have no measurable frequency drift despite the high frequency range. These pulsations
are not well cross-correlated at individual frequencies, and they have zero polarization.
Furthermore, the equality between LCP and RCP polarization during the whole event might be
caused by strong depolarization processes, for example, by a~strong plasma turbulence in the
radio source. This turbulent plasma scenario is supported by the facts of no radio
data correlation (Fig.~2, panel~$a$) and no characteristic dominant period in the period
range 0.07--1.49\,s (Fig.~1, bottom panel).

We tried to find the most likely possible interpretation of these pulsations in the very
complex topology of the active region.

Naturally, the microwave emission observed in the solar flares is frequently linked with
the gyrosynchrotron emission of mildly relativistic electrons. This incoherent radio
emission is shown in panel~$k$ (Fig.~2) at frequencies $>$~6\,GHz (radio continuum
without frequency drift). The other types of the bursts in the left panels (Fig.2) need
another explanation. We found that the radio fluxes in panels $a$~and~$c$ (Fig.2)
correspond to the tadpole patterns shown in Figs. 3~and~5. We interpret these
patterns found in the wavelet spectra of radio emission fluxes as a~signatures of the
fast sausage magnetoacoustic wave trains moving along a~dense flare loop (waveguide) and
passing through the radio source in the loop.

These magnetoacoustic wave trains are characterized by the steady periodical tail and the
quasi-periodical head that can be visualized with the~help of their wavelet power spectra.
This characteristic wavelet tadpole pattern consists of the long-period spectral
components (tadpole tail with a~characteristic period~$P$) propagating faster than the
medium and short-period ones (tadpole head). Numerical simulations of these tadpole
patterns can be found in Nakariakov et al. (2004) and M\'esz\'arosov\'a et al.
(2014), for instance. Observed fast sausage magnetoacoustic waves with these tadpoles were presented by
Katsiyannis et al. (2003) and M\'esz\'arosov\'a et al. (2009a, b; 2011a, b; 2013).

We detected two cases with magnetoacoustic waves (Figs.~3 and~5) and tried to use
a~model based on MHD oscillations (e.g., Roberts et al. 1984, Nakariakov~\&~Melnikov
2009). In this model the period of oscillations $\tau$ is proportional to the Alfv\'en
transit time through the source: $\tau = 2.6(a/v_A)$, where $a$ is the small radius of
the cylindrical loop. Considering the Alv\'en velocity $v_A$~=~1610\,km\,s$^{-1}$ and
$\tau$~=~0.7\,s (Fig.~3), we obtained a very low value of $a$~=~433.5\,km for the radio
source. It is doubtful whether this can produce our broadband pulsations. Only highly
coherent plasma processes (when the brightness temperature is $T_b = 10^{14}-10^{16}$\,K)
can solve this problem. Moreover, this model is commonly applied for pulsations with
a~period greater than 1\,s (Nakariakov~\&~Melnikov 2009).

Below, we search for other arguments to explain the magnetoacoustic waves in the
active region. Therefore we discuss the~topology of this region in more
detail.

The expanding loop (Fig.~9) consisted of some temporal fine sub-structures observed, for example,
by SDO/AIA in channels 131\,\AA~channel during 9:34:33--9:35:21\,UT, which are shown in Fig.~11
(processed with the NAFE method). The position of the radio source is marked by an arrow in all
panels. While at 9:34:33\,UT the expanding loop is very compact and simple at the site
of the radio source, the later situation is different. Immediately before the radio
source activity (9:34:45\,UT), two substructures seem to be crossing in the~location of the
radio source. During the time of the radio source activity (9:34:57 and 9:35:09\,UT), these
individual sub-structures become increasingly complex and twisted (panels
$b$--$c$ in Fig.~8 also show twisted magnetic field lines in this area). Finally, a~new
sub-structure can be seen at 9:35:21\,UT (see the arrow in Fig.~11). Thus, we can
postulate the~hypothesis that (some of) these temporal sub-structures might be reconnected
and this process could be the~reason for the perturbations (bi-directional outflows)
detected in the time-slit diagram (panels~$b$--$c$ of Fig.~7). (We note that a~similar
time-slit diagram, but without the~separation according to individual characteristic
temporal structures, was presented in Sim\~{o}es at al. (2015) and interpreted as
the~result of reconnection). On the other hand, the actual data cadence and spatial
resolution of the available observations as well as the absence of the z-coordinates of the
event details do not allow us to make an unambiguous judgement.

The~question remains why only one of the two loops was expanding. For this purpose we searched
for the loop footpoint positions that might be connected with flare ribbons observed by
both the SDO/AIA (panels~$b$ and $d$ in Fig.~10) and the STEREO/SECCHI-EUVI
(panels~$e$--$h$ in Fig.~10) instruments. We found four such footpoints, which are marked by crosses
A--D in Figs.~9 and~10. Footpoints~A and~B are clearly separated, both of them are
connected with one of flare ribbon, and they were detected with~similar radial heights
(1.4 and 1.5\,Mm), that is, they are located in the solar chromosphere. The situation of footpoints~C and~D
is more complex. While footpoint~C is connected with the most active part of the
ribbon (of the highest EUV intensity), footpoint~D was localized slightly aside of
that ribbon part (see panel~$b$, Fig.~10). Footpoints~C and~D were detected instead at a
coronal radial hight of 3.5 Mm and a chromospheric hight of 2.3 Mm, respectively. It
seems that this very complex situation, which is connected with the ribbon marked by arrow~(2)
in Fig.~10 and footpoint~C, might cause the loop expansion. The non-expanding loop
seems to be connected instead with footpoint~D, which is localized next to the most active part of
the ribbon.

The flows detected in panels $b$--$c$ of Fig.~7 might be linked
with the magnetic reconnection outflows. We can see the flows propagating in both
direction along the slit (vertical orange solid line in Fig.~6). The 131 and
94\,\AA~flows start from about the location of the radio source center at 5700\,GHz (horizontal
solid line in Fig.~7 for 131\,\AA). Moreover, the flows in panel $b)$ are limited to the
vicinity of the radio source. (We note that the EUV time-slit diagram with
reconnection inflows was studied in Yokoyama 2001). Therefore, some link between the EUV
bidirectional flows and the reconnection outflows might be possible.

We suggested an explanation of the solar event observed by the
different instruments at different spectral ranges. To confirm the
proposed complex scenario, observational data with a~higher spatial resolution are necessary.

\begin{acknowledgements}
We thank the referee for very useful comments that improved this paper. Data are courtesy
of the SDO/AIA, SDO/HMI, RHESSI and STEREO/SECCHI teams. This research was supported by
grants P209/12/0103 (GA CR), P209/10/1680 (GA CR), the research project RVO:67985815 of
the Astronomical Institute AS, the Marie Curie PIRSES-GA-2011-295272 RadioSun project,
the Slovak Research and Development Agency under the contract No. APVV-0816-11 and by the
Science Grant Agency project VEGA 2/0004/16. Help of the Bilateral Mobility Project
SAV-16-03 of the SAS and CAS is acknowledged. This article was created by the realization
of the project ITMS No. 26220120009, based on the supporting operational Research and
development program financed from the European Regional Development. I.M. acknowledges
support from grants 15-02-03835~a, 15-32-20504~mol~a~ved, and 16-32-00315~mol~a. The
wavelet analysis was performed with software based on tools provided by C.~Torrence and
G.~P.~Compo at \texttt{http://paos.colorado.edu/research/wavelets}.
\end{acknowledgements}

\end{document}